\documentclass[12pt]{iopart}

\usepackage{graphicx}
\usepackage[english,francais]{babel}

\def\ep{\varepsilon}

\def\dfrac{\displaystyle\frac}

\begin{document}
\selectlanguage{english}
\title[Focussing light in a bi-anisotropic slab with negatively refracting materials]{Focussing light in a bi-anisotropic slab with negatively refracting materials}

\author{Yan Liu$^1$, Sebastien Guenneau$^1$, Boris Gralak$^1$ and S. Anantha Ramakrishna$^2$}
\address{$^1$ Institut Fresnel, CNRS, Aix-Marseille Universit\'e, Ecole centrale Marseille, Campus de
Saint-J\'er\^ome, 13013 Marseille, France}
\address{$^2$ Department of Physics, Indian Institute of Technology, Kanpur 208016, India}

\begin{abstract}
We investigate the electromagnetic response of a pair of complementary bi-anisotropic media, which
consist of a medium with positive refractive index ($+\ep$, $+\mu$, $+\xi$) and a medium with negative
refractive index($-\ep$, $-\mu$, $-\xi$). We show that this idealized system has peculiar imaging properties
in that it reproduces images of a source, in principle, with unlimited resolution. We then consider an infinite array of line
sources regularly spaced in a one-dimensional photonic crystal (PC) consisting of 2n-layers of bi-anisotropic complementary media.
Using coordinate transformation, we map this system into 2D corner chiral lenses of 2n heterogeneous anisotropic
complementary media sharing a vertex, within which light
circles around closed trajectories. Alternatively, one can consider corner lenses with homogeneous isotropic media
and map them onto one dimensional PCs with heterogeneous bi-anisotropic layers. Interestingly, such complementary media
are described by scalar, or matrix valued, sign-shifting parameters,
which satisfy a generalized lens theorem, which can be derived using Fourier series solutions of the Maxwell's equations (in the former
case),
or from space-time symmetry arguments (in the latter case). Also of interest are 2D periodic checkerboards alternating rectangular cells of complementary media
which are such that one point source in one cell gives rise to an infinite set of images with an image in every other cell.
Such checkerboards can be mapped into a class of three-dimensional corner lenses of complementary bi-anisotropic media.

\end{abstract}
\maketitle

\section{Introduction}
\begin{figure}
    \centering
    \includegraphics[scale=0.6]{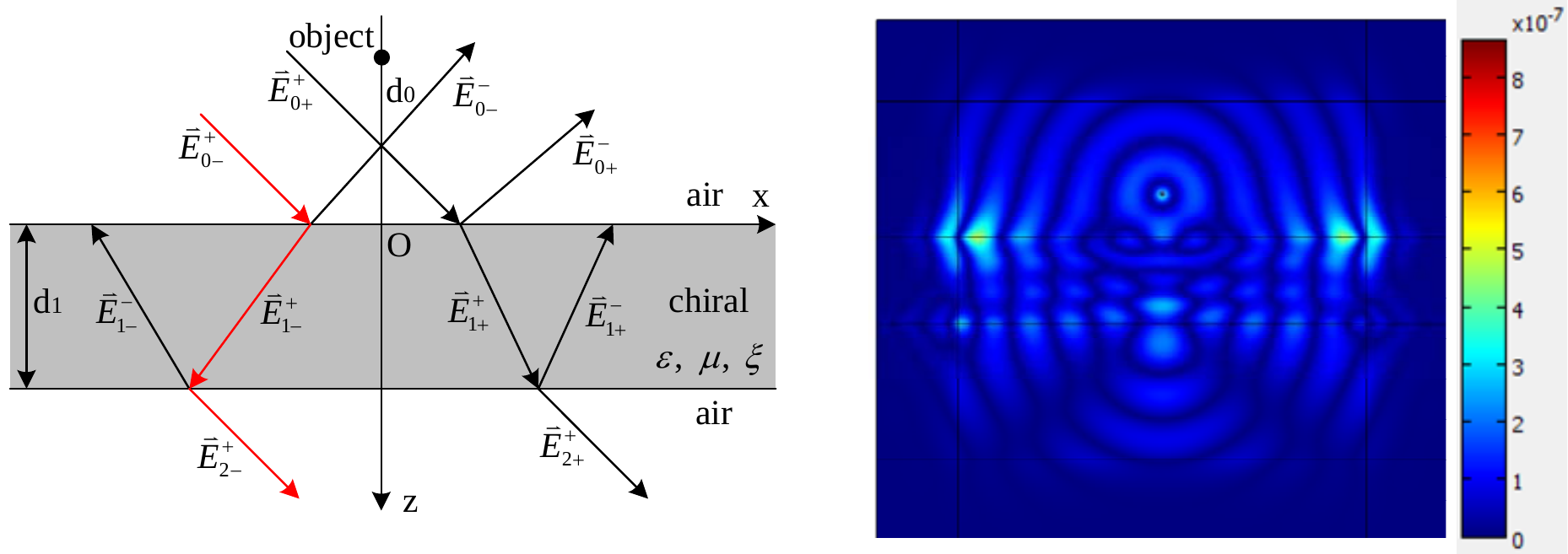}
    \caption{(a) Configuration for the focusing of a chiral slab for an object in air. Subscripts '+' and '-' of the electric field spectra indicate the left- and right-going wave streams, respectively. (b) Plot of $\sqrt{e^2+h^2}$ for a time harmonic electric line source radiating in the chiral slab by Finite element method.}
   \label{figure1}
\end{figure}

Jin and He recently proposed to use a slab of chiral medium in order to achieve a refocussing effect via negative refraction \cite{he}.
These authors noted that at the interface between a conventional medium with permittivity $\varepsilon_0$ and permeability
$\mu_0$, and a chiral medium with permittivity $\varepsilon_1$ and permeability $\mu_1$, and chirality $\kappa$, the
refractive indices for circularly polarized waves associated with wavenumbers $k_\pm=\omega(\kappa\sqrt{\mu_0\varepsilon_0}\pm\sqrt{\mu_1\varepsilon_1})$
are $n_\pm=\sqrt{(\mu_1\varepsilon_1)/(\mu_0\varepsilon_0)}\pm\kappa$. Indeed, when a wave is obliquely incident upon such an interface, and if $\kappa\sqrt{\varepsilon_0\mu_0}>\sqrt{\varepsilon_1\mu_1}$, then $n_{-}$ is clearly negative,
so that circularly polarized waves with wavenumber $k_{-}$ undergo some form of negative refraction, as shown in figure \ref{figure1} (rays in red color), which is the earlier result of Jin and He\cite{he}. Wiltshire and Pendry also proved a strong chiral in chiral Swiss rolls, which can be larger than either the permittivity or the permeability \cite{Wiltshire}. However, it is also legitimate to ask whether one can achieve similar focussing effects with more general bianisotropic media with sign shifting
in permittivity, permeability and bianisotropy. Furthermore, one can also consider heterogeneous anisotropic parameters. One way to answer such questions is to consider
complementary media which cancel out optical space \cite{pendry}, and to extend their analysis to chiral complementary media. Using the powerful tools of transformation
optics, one can also extend the design of generalized perfect lenses and corners to chiral and bianisotropic media. Finite element methods shall be used to numerically explore such optical systems. These analytical solutions of the singular systems considered by us represent important benchmarks for the validation of non-trivial numerical calculations.
\section{A Generalized Lens Theorem for chiral media}

\subsection{Proof via Fourier analysis}
The original 'perfect lens' as introduced by Victor Veselago[1968] \cite{veselago} is defined by a slab of material with $\varepsilon=-1$, $\mu=-1$, however, a much more general condition as proposed by Pendry and Ramakrishna in 2003 \cite{pendry} is the system described by,
\begin{equation}
\begin{array}{cc}
  \varepsilon_1=+\varepsilon(x,y), \quad \mu_1=+\mu(x,y), \quad -d<z<0  \\[2mm]
  \varepsilon_2=-\varepsilon(x,y), \quad \mu_2=-\mu(x,y), \quad 0<z<d  \\[2mm]
\end{array}
\end{equation}
Within this framework, a new theorem can be stated to prove the perfect lens: it is enough to point out that two complementary media have optical sum zero (the media in the region $-d<z<d$ behave as though they had zero thickness).

\begin{figure}
    \centering
    \includegraphics[scale=0.5]{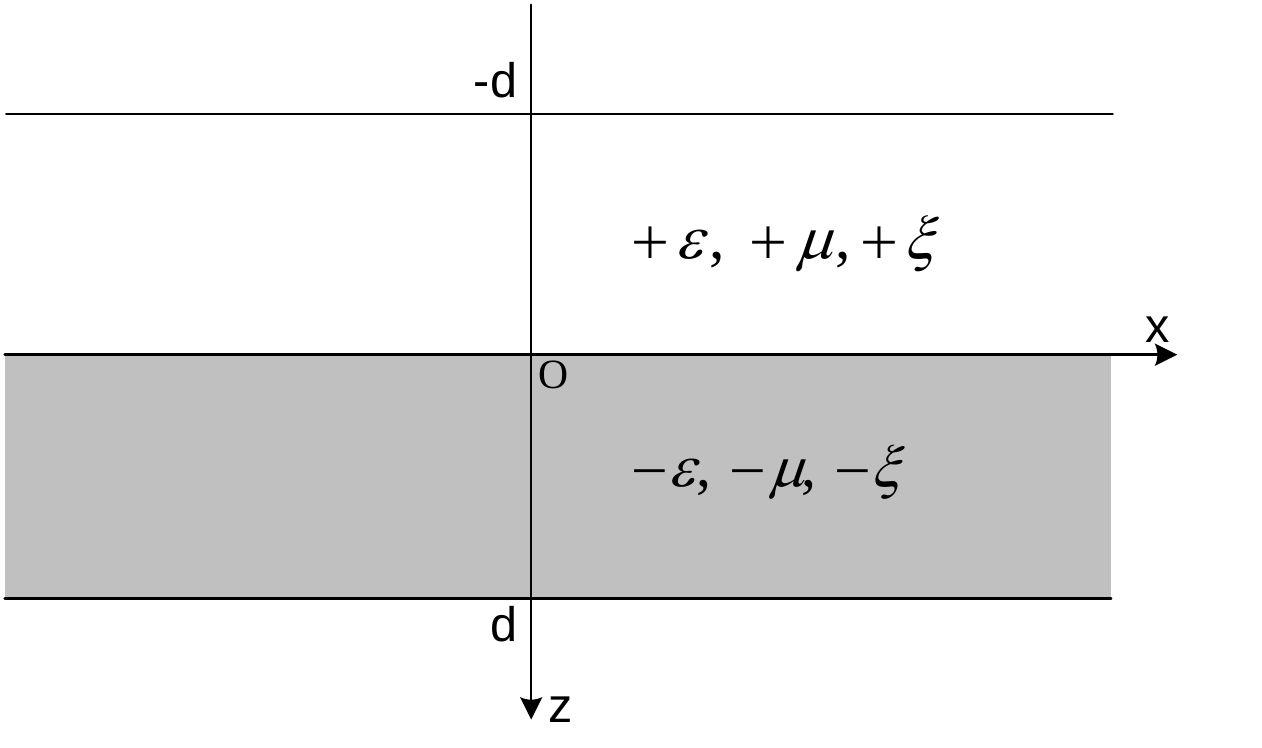}
    \caption{A complementary bianisotropic medium.}
    \label{figure2}
\end{figure}

However, when there is chirality in the complementary media, do they still work as a perfect lens? A new generalized lens
theorem should be considered. In what follows, we provide a positive answer for this kind of complementary bi-anisotropic media,

A system with two complementary bianisotropic media is shown in figure \ref{figure2},
\begin{equation}
\begin{array}{cc}
  \varepsilon_1=+\varepsilon(x,y), \quad \mu_1=+\mu(x,y), \quad \xi_1=+\xi(x,y), \quad -d<z<0 \\[2mm]
  \varepsilon_2=-\varepsilon(x,y), \quad \mu_2=-\mu(x,y), \quad \xi_2=-\xi(x,y), \quad 0<z<d \\[2mm]
\end{array}
\end{equation}
Let us assume constitutive relations in a bianisotropic medium as follows,
\begin{equation}
  {\bf D}=\ep {\bf E}+i\xi {\bf H}, \qquad {\bf B}=\mu {\bf H}-i\xi {\bf E}
\label{conequ}
\end{equation}
with $\xi$ the chirality parameter, instead of $\kappa$ in He \cite{he}.

Maxwell's equations take the form (known as Maxwell-Tellegen's equations),
\begin{equation}
\begin{array}{cc}
\nabla \times {\bf E}= \mu \dfrac{\partial{\bf H}}{\partial t}- i \xi \dfrac{\partial{\bf E}}{\partial t}, \nonumber\\[2mm]
\nabla \times {\bf H}=-\ep \dfrac{\partial{\bf E}}{\partial t}- i \xi \dfrac{\partial{\bf H}}{\partial t}
\end{array}
\label{max}
\end{equation}
Let us decompose the electric fields in Fourier series,
\begin{equation}
\fl \begin{array}{cc}
{\bf E_1}(x,y,z)=\exp(+ik_{1z}z)\sum\limits_{k_x,k_y}{\bf E_1}(k_x,k_y) \exp(ik_x x+ik_y y), \quad -d<z<0 \\[2mm]
{\bf E_2}(x,y,z)=\exp(+ik_{2z}z)\sum\limits_{k_x,k_y}{\bf E_2}(k_x,k_y) \exp(ik_x x+ik_y y), \quad 0<z<d  \\[2mm]
\end{array}
\label{fourier}
\end{equation}
The same holds true for the magnetic fields.
These fields should satisfy some boundary conditions: The tangential components of both electric and magnetic fields are continuous across the interface at $z=0$.

Plugging the Fourier series (\ref{fourier}) for the electric and magnetic fields into (\ref{max}), and using the frequency domain notation $\partial/\partial t = i \omega$, we get,
\begin{eqnarray}
  k_{1z}{\hat z} &\times \big[ E_{1x}(k_x,k_y) {\hat x}+ E_{1y}(k_x,k_y) {\hat y}\big]
  +(k_x{\hat x}+k_y{\hat y}) \times E_{1z}(k_x,k_y) {\hat z} \nonumber \\[2mm]
  &=\omega \sum_{k'_x,k'_y} \mu_1(k_x,k_y,k'_x,k'_y)\big[ H_{1x}(k'_x,k'_y) {\hat x}+H_{1y}(k'_x,k'_y) {\hat y}\big] \nonumber \\
  &-i\omega \sum_{k'_x,k'_y} \xi_1(k_x,k_y,k'_x,k'_y)\big[ E_{1x}(k'_x,k'_y) {\hat x}+E_{1y}(k'_x,k'_y) {\hat y}\big] \nonumber \\
 (k_x{\hat x}&+k_y{\hat y})\times \big[ E_{1x}(k_x,k_y) {\hat x}+ E_{1y}(k_x,k_y) {\hat y}\big] \nonumber\\[2mm]
  &=\omega \sum_{k'_x,k'_y} \mu_1(k_x,k_y,k'_x,k'_y) H_{1z}(k_x,k_y,k'_x,k'_y) {\hat z} \nonumber \\
  &-i\omega \sum_{k'_x,k'_y} \xi_1(k_x,k_y,k'_x,k'_y) E_{1z}(k_x,k_y,k'_x,k'_y) {\hat z} \nonumber \\
  k_{1z}{\hat z} & \times \big[ H_{1x}(k_x,k_y) {\hat x}+ H_{1y}(k_x,k_y) {\hat y}\big]
  +(k_x{\hat x}+k_y{\hat y}) \times H_{1z}(k_x,k_y) {\hat z} \nonumber \\[2mm]
  &=- \sum_{k'_x,k'_y} \varepsilon_1(k_x,k_y,k'_x,k'_y)\big[ E_{1x}(k'_x,k'_y) {\hat x}+E_{1y}(k'_x,k'_y) {\hat y}\big] \nonumber \\
  &-i\omega \sum_{k'_x,k'_y} \xi_1(k_x,k_y,k'_x,k'_y)\big[ H_{1x}(k'_x,k'_y) {\hat x}+H_{1y}(k'_x,k'_y) {\hat y}\big] \nonumber \\
  (k_x{\hat x}&+k_y{\hat y})\times \big[ H_{1x}(k_x,k_y) {\hat x}+ H_{1y}(k_x,k_y) {\hat y}\big] \nonumber \\[2mm]
  &=-\omega \sum_{k'_x,k'_y} \varepsilon_1(k_x,k_y,k'_x,k'_y) H_{1z}(k_x,k_y,k'_x,k'_y) {\hat z} \nonumber \\
  &-i\omega \sum_{k'_x,k'_y} \xi_1(k_x,k_y,k'_x,k'_y) H_{1z}(k_x,k_y,k'_x,k'_y) {\hat z} \nonumber \\
\label{ex}
\end{eqnarray}
If we now substitute,
\begin{eqnarray}
  \fl E_{2x}(k_x,k_y)=E_{1x}(k_x,k_y), \quad E_{2y}(k_x,k_y)=E_{1y}(k_x,k_y),\quad E_{2z}(k_x,k_y)=-E_{1z}(k_x,k_y), \nonumber \\[2mm]
  \fl  H_{2x}(k_x,k_y)=H_{1x}(k_x,k_y), \quad H_{2y}(k_x,k_y)=H_{1y}(k_x,k_y),\quad H_{2z}(k_x,k_y)=-H_{1z}(k_x,k_y), \nonumber \\[2mm]
  \fl  \varepsilon_2(k_x,k_y,k'_x,k'_y)=-\varepsilon_1(k_x,k_y,k'_x,k'_y), \quad
   \mu_2(k_x,k_y,k'_x,k'_y)=-\mu_1(k_x,k_y,k'_x,k'_y), \nonumber \\[2mm]
  \fl  \xi_2(k_x,k_y,k'_x,k'_y)=-\xi_1(k_x,k_y,k'_x,k'_y), \quad k_{2z}=-k_{1z}
\end{eqnarray}
into (\ref{ex}), we obtain,
\begin{eqnarray}
  -k_{2z}{\hat z} &\times \big[ E_{2x}(k_x,k_y) {\hat x}+E_{2y}(k_x,k_y) {\hat y}\big]
  -(k_x{\hat x}+k_y{\hat y}) \times E_{2z}(k_x,k_y) {\hat z} \nonumber \\[2mm]
  &=-\omega \sum_{k'_x,k'_y} \mu_2(k_x,k_y,k'_x,k'_y)\big[ H_{2x}(k'_x,k'_y) {\hat x}+H_{2y}(k'_x,k'_y) {\hat y}\big] \nonumber \\
  &+i\omega \sum_{k'_x,k'_y} \xi_2(k_x,k_y,k'_x,k'_y)\big[ E_{2x}(k'_x,k'_y) {\hat x}+E_{2y}(k'_x,k'_y) {\hat y}\big] \nonumber \\
  (k_x{\hat x}&+k_y{\hat y}) \times \big[ E_{2x}(k_x,k_y) {\hat x}+ E_{2y}(k_x,k_y) {\hat y}\big] \nonumber \\[2mm]
  &=\omega \sum_{k'_x,k'_y} \mu_2(k_x,k_y,k'_x,k'_y) H_{2z}(k_x,k_y,k'_x,k'_y) {\hat z} \nonumber \\
  &-i\omega \sum_{k'_x,k'_y} \xi_2(k_x,k_y,k'_x,k'_y) E_{2z}(k_x,k_y,k'_x,k'_y) {\hat z} \nonumber\\
  -k_{2z}{\hat z} & \times \big[ H_{2x}(k_x,k_y) {\hat x}+ H_{2y}(k_x,k_y) {\hat y}\big]
  -(k_x{\hat x}+k_y{\hat y}) \times H_{2z}(k_x,k_y) {\hat z} \nonumber \\[2mm]
  &=\omega \sum_{k'_x,k'_y} \varepsilon_2(k_x,k_y,k'_x,k'_y)\big[ E_{2x}(k'_x,k'_y) {\hat x}+E_{2y}(k'_x,k'_y) {\hat y}\big] \nonumber \\
  &+i\omega \sum_{k'_x,k'_y} \xi_2(k_x,k_y,k'_x,k'_y)\big[ H_{2x}(k'_x,k'_y) {\hat x}+H_{2y}(k'_x,k'_y) {\hat y}\big] \nonumber \\
  (k_x{\hat x}&+k_y{\hat y}) \times \big[ H_{2x}(k_x,k_y) {\hat x}+ H_{2y}(k_x,k_y) {\hat y}\big] \nonumber \\[2mm]
  &=-\omega \sum_{k'_x,k'_y} \varepsilon_2(k_x,k_y,k'_x,k'_y) H_{2z}(k_x,k_y,k'_x,k'_y) {\hat z}\nonumber \\
  &-i\omega \sum_{k'_x,k'_y} \xi_2(k_x,k_y,k'_x,k'_y) H_{2z}(k_x,k_y,k'_x,k'_y) {\hat z} \nonumber \\
\end{eqnarray}
These equations can be rearranged as,
\begin{eqnarray}
  k_{2z}{\hat z} &\times \big[ E_{2x}(k_x,k_y) {\hat x}+E_{2y}(k_x,k_y) {\hat y}\big]
  +(k_x{\hat x}+k_y{\hat y}) \times E_{2z}(k_x,k_y) {\hat z} \nonumber \\[2mm]
  &=\omega  \sum_{k'_x,k'_y} \mu_2(k_x,k_y,k'_x,k'_y)\big[ H_{2x}(k'_x,k'_y) {\hat x}+H_{2y}(k'_x,k'_y) {\hat y}\big] \nonumber \\
  &-i\omega \sum_{k'_x,k'_y} \xi_2(k_x,k_y,k'_x,k'_y)\big[ E_{2x}(k'_x,k'_y) {\hat x}+E_{2y}(k'_x,k'_y) {\hat y}\big] \nonumber \\
  (k_x{\hat x} &+k_y{\hat y}) \times \big[ E_{2x}(k_x,k_y) {\hat x}+ E_{2y}(k_x,k_y) {\hat y}\big] \nonumber \\[2mm]
  &=\omega  \sum_{k'_x,k'_y} \mu_2(k_x,k_y,k'_x,k'_y) H_{2z}(k_x,k_y,k'_x,k'_y) {\hat z} \nonumber \\
  &-i\omega \sum_{k'_x,k'_y} \xi_2(k_x,k_y,k'_x,k'_y) E_{2z}(k_x,k_y,k'_x,k'_y) {\hat z} \nonumber \\
  k_{2z}{\hat z} & \times \big[ H_{2x}(k_x,k_y) {\hat x}+ H_{2y}(k_x,k_y) {\hat y}\big]
  +(k_x{\hat x}+k_y{\hat y}) \times H_{2z}(k_x,k_y) {\hat z} \nonumber \\[2mm]
  &=-\omega \sum \varepsilon_2(k_x,k_y,k'_x,k'_y)\big[ E_{2x}(k'_x,k'_y) {\hat x}+E_{2y}(k'_x,k'_y) {\hat y}\big] \nonumber \\
  &-i\omega \sum_{k'_x,k'_y} \xi_2(k_x,k_y,k'_x,k'_y)\big[ H_{2x}(k'_x,k'_y) {\hat x}+H_{2y}(k'_x,k'_y) {\hat y}\big] \nonumber \\
  (k_x{\hat x}& +k_y{\hat y}) \times \big[ H_{2x}(k_x,k_y) {\hat x}+ H_{2y}(k_x,k_y) {\hat y}\big] \nonumber \\[2mm]
  &=-\omega \sum_{k'_x,k'_y} \varepsilon_2(k_x,k_y,k'_x,k'_y) H_{2z}(k_x,k_y,k'_x,k'_y) {\hat z}\nonumber \\
  &-i\omega \sum_{k'_x,k'_y} \xi_2(k_x,k_y,k'_x,k'_y) H_{2z}(k_x,k_y,k'_x,k'_y) {\hat z} \nonumber \\
  \label{ex2}
\end{eqnarray}
Equations (\ref{ex}) and (\ref{ex2}) hold for the fields in chiral media with parameters $\ep_1, \mu_1, \xi_1$ and $\ep_2, \mu_2, \xi_2$ respectively.
The fields in the two regions match across the interface due to the continuous boundary condition, while there is a sign-shifting between components along $z$ direction, hence,
\begin{eqnarray}
  &{\bf E}(x,y,z=+d)={\bf E}(x,y)\exp{(-ik_z d)} \nonumber \\[2mm]
  &={\bf E}(x,y)\exp{(ik_z(-d))}={\bf E}(x,y,z=-d), \,\, d>0.
\label{com}
\end{eqnarray}
(\ref{com}) holds for each $k_z$ and hence the sums over $k_z$, the same result holds for the magnetic fields. We would like now to extend these results to anisotropic
tensors of permittivity, permeability and chirality. This can be done using the same kind of techniques as above, but computations become tedious. We therefore propose another,
more subtle, but more concise, derivation of the generalized lens theorem using space-time symmetries.
\subsection{Proof via space-time symmetries}
Rewrite Maxwell-Tellegen's equations (\ref{max}) of a bianisotropic medium as,
\begin{equation}
\begin{array}{cc}
  \nabla\times \,{\bf E} + i \dfrac{\partial \,{(\xi\bf E})}{\partial t} =\dfrac{\partial (\mu {\bf H})}{\partial t} \\[3mm]
  \nabla\times \,{\bf H} + i  \dfrac{\partial \,(\xi {\bf H})}{\partial t} =- \dfrac{\partial (\ep {\bf E})}{\partial t}
\label{max2}
\end{array}
\end{equation}
Let us list some transformations which leave the equation (\ref{max2}) invariant (see \cite{NJP2005} for the case when $\xi=0$):\\
{\bf S1} (Generalized conformal invariance): ${\bf E}\rightarrow \mathcal{A}{\bf E}$, ${\bf H}\rightarrow \mathcal{A}{\bf H}$,
${\mu^{-1}}\rightarrow \mathcal{A}{\mu^{-1}} \mathcal{A}^{-1}$, ${\varepsilon}\rightarrow \mathcal{A}{\varepsilon} \mathcal{A}^{-1}$,
${\xi}\rightarrow \mathcal{A}{\xi} \mathcal{A}^{-1}$, where $\mathcal{A}$ is invertible and an element of ${\emph {GL}}_3(\textbf{R})$
(a group of $3\times3$ linear operators). \\
{\bf S2} (Generalized duality): ${\bf E} \rightarrow -{\bf H}$, ${\bf H} \rightarrow -{\bf E}$, (iff $\mu=\ep$).  \\
{\bf S3} (Parity invariance): ${\bf r} \rightarrow -{\bf r}$, where ${\bf r}=[x,y,z]$, Maxwell-Tellegen's equations in (\ref{max2}) turn out to be,
\begin{equation}
\begin{array}{cc}
  -\nabla\times \,{\bf E} + i \dfrac{\partial \,{(\xi\bf E})}{\partial t} =-\dfrac{\partial (\mu {\bf H})}{\partial t} \\[3mm]
  -\nabla\times \,{\bf H} + i  \dfrac{\partial \,(\xi {\bf H})}{\partial t} = -\dfrac{\partial (\ep {\bf E})}{\partial t}
\label{max2-1}
\end{array}
\end{equation}
in order to preserve equation (\ref{max2}), ${\bf E} \rightarrow -{\bf E}$, ${\bf H} \rightarrow -{\bf H}$, $\ep \rightarrow -\,\ep$, $\mu \rightarrow -\,\mu$, $\xi \rightarrow -\,\xi$ should be chosen to cancel out the minus introduced by $-{\bf r}$.  \\
{\bf S4} (Time reversal): $t \rightarrow -t$, then ${\bf E} \rightarrow -{\bf E}$ (or ${\bf H} \rightarrow -{\bf H}$) and $\xi \rightarrow -\,\xi$ (similar idea as Parity invariance).  \\
{\bf S5} Any additional space-time symmetries. \\
The combination of any of these symmetries is again a symmetry of the system of equations. In other words, if the fields in a particular region of space can be
mapped on to another region of space through the symmetry transformations S1-S6 while preserving the respective boundary conditions, then the transformed fields solve the field equations whenever the original fields do. This law will also be proved in the following part with coordinates transformation.

\begin{figure}
    \centering
    \includegraphics[scale=0.7]{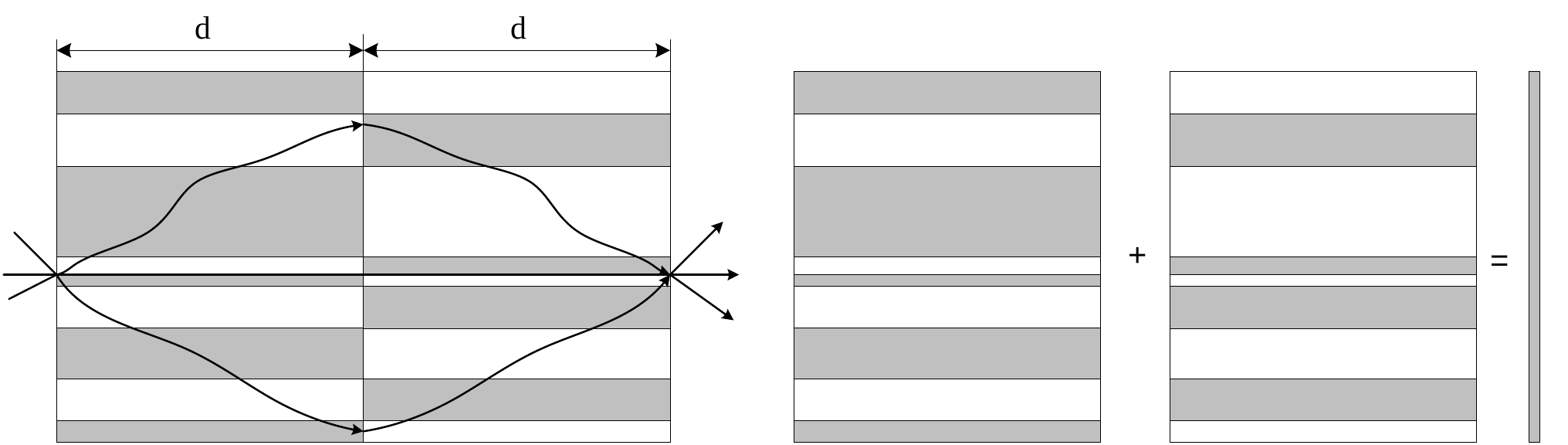}
    \caption{An alternative pair of complementary chiral media, each cancelling the effect of the other, that means
    complementary halves sum to zero.}
    \label{figure3}
\end{figure}

Let us consider a homogeneous slab in figure \ref{figure2} with permittivity, permeability and chirality tensors in the region of $-d<z<0$,
\begin{equation}
\begin{array}{cc}
  v_1=\left[
  \begin{array}{ccc}
    v_{xx} & v_{xy} & v_{xz} \\
    v_{yx} & v_{yy} & v_{yz} \\
    v_{zx} & v_{zy} & v_{zz} \\
  \end{array}
  \right], \quad v=\ep,\, \mu,\, \xi
\end{array}
\label{comp}
\end{equation}
For a source at the interface and propagation along the $z$ direction, we use the symmetry operations S3, followed by S1 with,
\begin{equation}
  \mathcal{A}=\left[
  \begin{array}{ccc}
    -1 & 0 & 0 \\
    0 & -1 & 0\\
    0 & 0 & 1
  \end{array}\right]
\end{equation}
We will call this sequence of operations a mirror operation. The choice of $\mathcal{A}$ preserves the continuity of ${\bf E}_\parallel$ and $ {\bf H}_\parallel$ across the boundary. Then the complementary medium of $0<z<d$ is,
\begin{equation}
\begin{array}{cc}
  v_2=\left[
  \begin{array}{ccc}
    -v_{xx} & -v_{xy} & v_{xz} \\
    -v_{yx} & -v_{yy} & v_{yz} \\
    v_{zx} & v_{zy} & -v_{zz} \\
  \end{array}
  \right], \quad v=\ep,\, \mu,\, \xi
\end{array}
\end{equation}
Similar to the complementary theory proposed by Pendry and Ramakrishna, a schematic diagram for the bianisotropic complementary media is indicated in figure \ref{figure3},
two complementary bianisotropic media have an optical sum of zero. The optical response on the two sides of a structure with complementary media can be calculated by cutting out the complementary media and closing the gap between the two sides.

The two derivations of the generalized lens theorem are underpinned by the invariance of the Maxwell's equations under geometric changes. If one wants to study other classes of generalized
lenses, say in polar or spherical coordinates, some more elaborate mathematical tools are needed, as we shall now see.
\section{Transformation optics applied to the design of corner lenses for bianisotropic media}

\subsection{General Coordinates Transformation}
\begin{figure}
    \centering
    \includegraphics[scale=0.7]{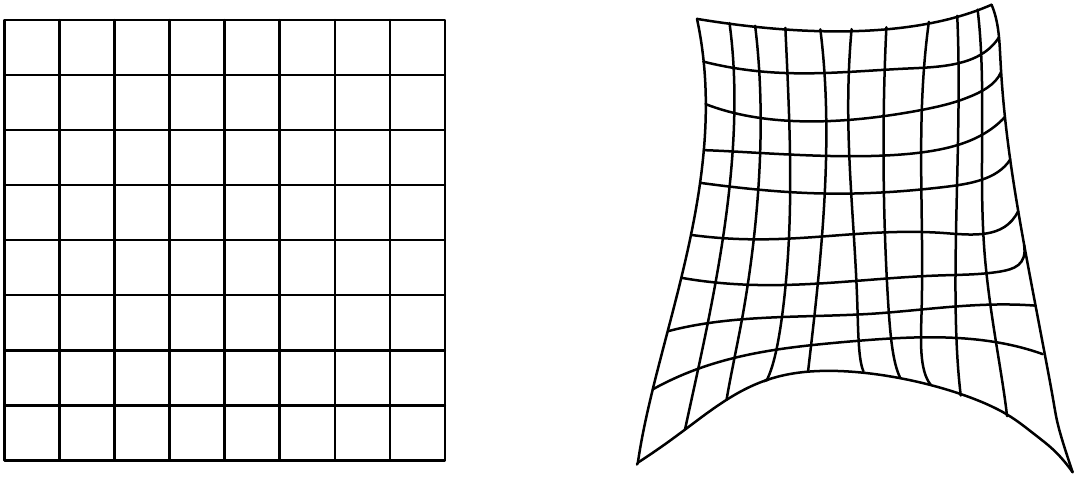}
    \caption{A simple cubic lattice in one coordinate system (left) maps into a distorted mesh in the other coordinate system(right).}
    \label{figure4}
\end{figure}

Considering a new coordinate system described by\cite{ward96},
\begin{equation}
  q_1(x,y,z),\quad q_2(x,y,z),\quad q_3(x,y,z)
\end{equation}
Lines of constant $q_2$, $q_3$ define the generalized $q_1$ axes, and so on. It should be noted that we can produce  a desired mesh by choosing the proper coordinates transformation. For example, if we define a set of points by equal increments along the new coordinates $q_1$, $q_2$ and $q_3$ axes, these will make the mesh in the original coordinates system look like distorted as shown in figure \ref{figure4}.

Since Maxwell-Tellegen's equations have the forms of (\ref{max}) in a Cartesian coordinate system, while for a new coordinate system of $q_1$, $q_2$, $q_3$ (which can be non-orthogonal), one may wonder what will happen to Maxwell-Tellegen's equations, i.e. will they preserve their form like,
\begin{equation}
\begin{array}{ll}
\nabla_q \times {\bf {\hat E}}= {\hat \mu} \dfrac{\partial{\bf{\hat H}}}{\partial t}- i{\hat \xi} \dfrac{\partial{\bf {\hat E}}}{\partial t}, \\[2mm]
\nabla_q \times {\bf {\hat H}}=-{\hat \ep} \dfrac{\partial{\bf {\hat E}}}{\partial t}- i{\hat \xi} \dfrac{\partial{\bf {\hat H}}}{\partial t}
\end{array}
\label{newmaxwell}
\end{equation}
where $\hat {\bf E}$, $\hat {\bf H}$ are re-normalized electric and magnetic fields, and $\hat \ep$, $\hat \mu$ and $\hat \xi$ are in general tensors? While this has been discussed by Ward and Pendry for anisotropic systems \cite{ward96}, we present here a generalization of those results to fully bianisotropic systems.

The proof of the form invariance of the Maxwell-Tellegen's equations is detailed below:

First, we define three units vectors. Let ${\bf u_1}$, ${\bf u_2}$, ${\bf u_3}$ be along the axes of $q_1$, $q_2$, $q_3$. The transformation relation between $[x,y,z]$ and $[q_1,q_2,q_3]$ is obviously,
\begin{equation}
  \left[ \begin{array}{l}
  dx\\[2mm]
  dy\\[2mm]
  dz \end{array} \right]=
  \left[\begin{array}{lll}
  \dfrac{\partial x}{\partial q_1} & \dfrac{\partial x}{\partial q_2} & \dfrac{\partial x}{\partial q_3} \\[3mm]
  \dfrac{\partial y}{\partial q_1} & \dfrac{\partial y}{\partial q_2} & \dfrac{\partial y}{\partial q_3} \\[3mm]
  \dfrac{\partial z}{\partial q_1} & \dfrac{\partial z}{\partial q_2} & \dfrac{\partial z}{\partial q_3}
  \end{array}\right]
  \left[ \begin{array}{l}
  d{q_1}\\[2mm]
  d{q_2}\\[2mm]
  d{q_3} \end{array} \right]
\end{equation}
the length of a line element in the new coordinate system can be derived from
\begin{eqnarray}
  ds^2=dx^2+dy^2+dz^2&=Q_{11}d{q_1^2}+Q_{22}d{q_2^2}+Q_{33}d{q_3^2} \nonumber \\
  &+2Q_{12}d{q_1}d{q_2}+2Q_{13}d{q_1}d{q_3}+2Q_{23}d{q_2}d{q_3}
\end{eqnarray}
with
\begin{equation}
  Q_{ij}=\dfrac{\partial x}{\partial q_i}\dfrac{\partial x}{\partial q_j}+\dfrac{\partial y}{\partial q_i}\dfrac{\partial y}{\partial q_j}
  +\dfrac{\partial z}{\partial q_i}\dfrac{\partial z}{\partial q_j}
\end{equation}

Let us now note that the electric field ${\bf E}$ and magnetic field ${\bf H}$ can be expressed in terms of contravariant components,
\begin{equation}
 {\bf U} = U^1 {\bf u_1} + U^2 {\bf u_2} + U^3 {\bf u_3}, \quad U= E,\, H
 \label{contrav}
\end{equation}
and conversely, they can be expressed in terms of the covariant components through
\begin{equation}
  g^{-1} \left[ \begin{array}{c}
  U^1 \\
  U^2 \\
  U^3
  \end{array}\right] =\left[  \begin{array}{ccc}
  {\bf u_1} \cdot {\bf u_1} & {\bf u_1} \cdot {\bf u_2} & {\bf u_1} \cdot {\bf u_3} \\
  {\bf u_2} \cdot {\bf u_1} & {\bf u_2} \cdot {\bf u_2} & {\bf u_2} \cdot {\bf u_3} \\
  {\bf u_3} \cdot {\bf u_1} & {\bf u_3} \cdot {\bf u_2} & {\bf u_3} \cdot {\bf u_3}
  \end{array}\right] \left[ \begin{array}{c}
  U^1 \\
  U^2 \\
  U^3
  \end{array}\right] = \left[ \begin{array}{c}
  U_1 \\
  U_2 \\
  U_3
  \end{array}\right]
\label{trans}
\end{equation}
with $g$ a 3*3 matrix and
\begin{equation}
  U_1 = {\bf U} \cdot {\bf u_1}, \quad  U_2 = {\bf U} \cdot {\bf u_2} ,\quad  U_3= {\bf U} \cdot {\bf u_3}.
\label{U}
\end{equation}
according to (\ref{trans}),
\begin{equation}
  U^i=\sum_{j=1}^{3} g^{ij} U_j
\label{con-co}
\end{equation}

\begin{figure}
    \centering
    \includegraphics[scale=0.7]{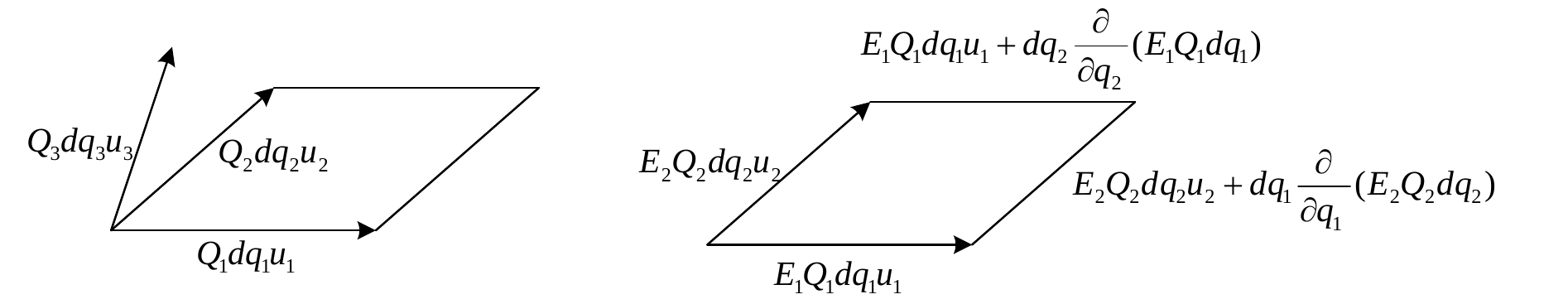}
    \caption{(a) Small element resembling a parallelepiped, (b) Integration path for finding $\nabla \times {\bf E}$.}
    \label{figure5}
\end{figure}

We take a small element which resembles a parallelepiped (left panel in figure \ref{figure5}), and consider the projection of $\nabla \times {\bf E}$ onto the normal to the ${\bf u_1}-{\bf u_2}$ plane by taking a line integral along the ${\bf u_1}-{\bf u_2}$ parallelogram (right panel of figure \ref{figure5}) \cite{ward96}, and invoking Stokes' theorem,
\begin{equation}
  {\int\int}_{S} \nabla \times {\bf E} \cdot d{\vec S} = \oint_{l} {\bf E} \cdot d{\vec l}
\label{stoke}
\end{equation}
where $S$ is the surface of ${\bf u_1}-{\bf u_2}$ plane and $l$ is the boundary around surface $S$.
Consequently,
\begin{equation}
  (\nabla \times {\bf E}) \cdot ({\bf u_1} \times {\bf u_2})Q_1dq_1 Q_2dq_2 = dq_1 \dfrac{\partial}{\partial q_1} (E_2 Q_2 dq_2) - dq_2 \dfrac{\partial}{\partial q_2} (E_1 Q_1 dq_1)
\label{Eint}
\end{equation}
on the other hand, the normalized electric field in new coordinate system is defined as,
\begin{equation}
  \hat E_i=Q_i E_i, \quad (i=1,2,3)
\label{newE}
\end{equation}
Hence equation (\ref{Eint}) can be simplified as,
\begin{equation}
 (\nabla \times {\bf E}) \cdot ({\bf u_1} \times {\bf u_2})Q_1 Q_2 = \dfrac{\partial \hat E_2}{\partial q_1}- \dfrac{\partial \hat E_1}{\partial q_2}= (\nabla_q \times \hat {\bf E})^3
\label{newEcurl}
\end{equation}
the term in the right-hand side of this equation is the third component of curl in the new coordinate system. Then applying (\ref{max}) to $\nabla \times {\bf E}$ in the left-hand side of equation (\ref{newEcurl}) we obtain
\begin{equation}
  (\nabla \times {\bf E}) \cdot ({\bf u_1} \times {\bf u_2})Q_1 Q_2 = \mu \dfrac{\partial {\bf H}}{\partial t}\cdot ({\bf u_1} \times {\bf u_2})Q_1 Q_2 - i\xi \dfrac{\partial {\bf E}}{\partial t}\cdot ({\bf u_1} \times {\bf u_2})Q_1 Q_2
\label{newEcurl2}
\end{equation}

Here it should be pointed out that the equation (15) in the paper of Ward and Pendry has a flaw for non-orthogonal coordinates system $q_1, q_2, q_3$. However, this does affect their result.
Let us now substitute (\ref{contrav}) combining with (\ref{con-co}) for both ${\bf E}$ and ${\bf H}$ to (\ref{newEcurl2}) and obtain
\begin{eqnarray}
 &(\nabla \times {\bf E}) \cdot ({\bf u_1} \times {\bf u_2})Q_1 Q_2 = \mu \dfrac{\partial {\bf H}}{\partial t}\cdot ({\bf u_1} \times {\bf u_2})Q_1 Q_2 - i\xi \dfrac{\partial {\bf E}}{\partial t}\cdot ({\bf u_1} \times {\bf u_2})Q_1 Q_2  \nonumber \\
  & = \mu \sum_{j=1}^{3} g^{3j} \dfrac{\partial H_j}{\partial t} {\bf u_3}\cdot ({\bf u_1} \times {\bf u_2})Q_1 Q_2 - i\xi \sum_{j=1}^{3} g^{3j} \dfrac{\partial E_j}{\partial t} {\bf u_3}\cdot ({\bf u_1} \times {\bf u_2})Q_1 Q_2
\label{nexmax}
\end{eqnarray}
define
\begin{equation}
  {\hat v}^{ij}= v g^{ij} |{\bf u_1}\cdot ({\bf u_2} \times {\bf u_3})|Q_1 Q_2 Q_3 (Q_i Q_j)^{-1},\quad v=\mu, \,\, \xi
\label{newpara}
\end{equation}
and
\begin{equation}
  {\hat U_j}=Q_j U_j
\end{equation}
hence (\ref{nexmax}) turns out to be
\begin{equation}
  (\nabla \times {\bf E}) \cdot ({\bf u_1} \times {\bf u_2})Q_1 Q_2 = \sum_{j=1}^{3} {\hat \mu}^{3j} \dfrac{\partial {\hat H_j}}{\partial t} - i\sum_{j=1}^{3} {\hat \xi}^{3j} \dfrac{\partial {\hat E_j}}{\partial t}
\end{equation}
comparing with (\ref{newEcurl}), we finally obtain
\begin{equation}
\nabla_q \times {\bf {\hat E}}= {\hat \mu} \dfrac{\partial{\bf{\hat H}}}{\partial t}- i{\hat \xi} \dfrac{\partial{\bf {\hat E}}}{\partial t},
\end{equation}
Similarly, due to the symmetry between $E$ and $H$ fields, we have
\begin{equation}
  \nabla_q \times {\bf {\hat H}}=-{\hat \ep} \dfrac{\partial{\bf {\hat E}}}{\partial t}- i{\hat \xi} \dfrac{\partial{\bf {\hat H}}}{\partial t}
\end{equation}
while $\hat \ep^{ij}$ also satisfies (\ref{newpara}).

These results indicate that the form of Maxwell-Tellegen's equations is preserved under a coordinates transformation, while the definitions of $\ep$, $\mu$ and $\xi$ are simply changed. In other words, no matter what is the coordinate system, we can always analyze the electromagnetic property of a structure through the Maxwell-Tellgen's equations with redefined parameters.

\subsection{1D PCs to 2D corner lenses}
In the previous section, we have analyzed a system with two complementary regions: the optical changes through these two regions is zero due to the fact that two regions are inverted mirror images of each other. In the following sections, we would like to give a further generalization of the cancelation principle combining with coordinates transformation.

\begin{figure}
    \centering
    \includegraphics[scale=0.7]{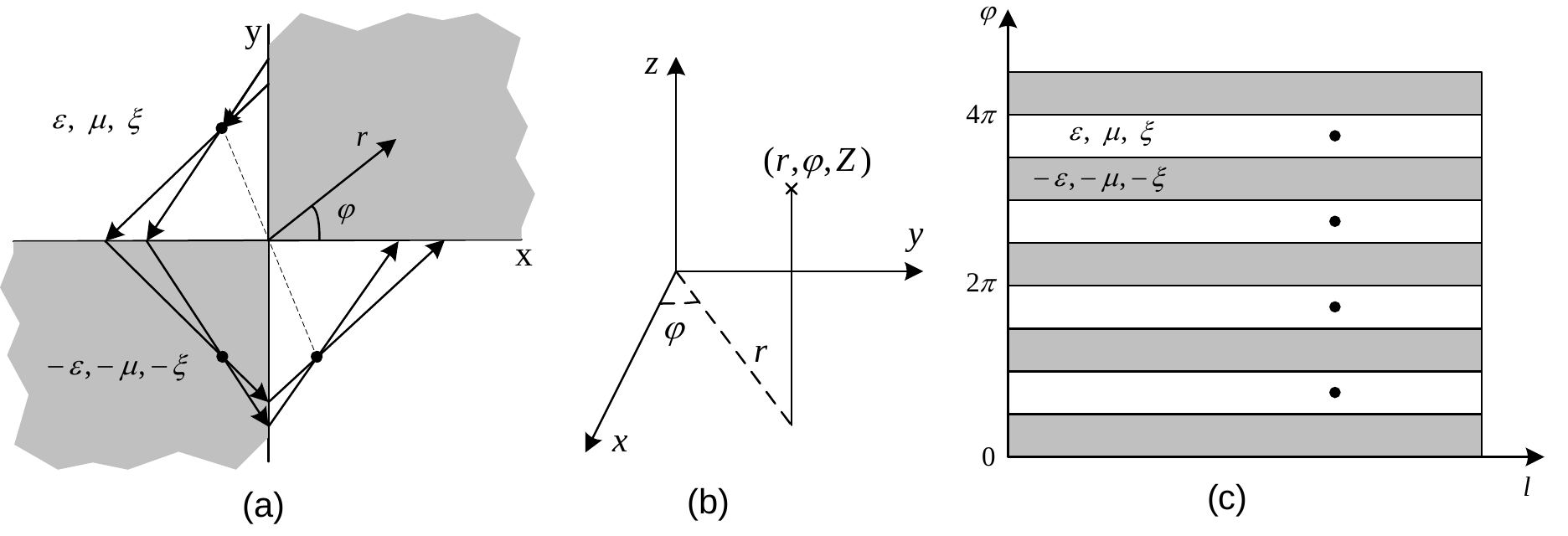}
    \caption{(a) A pair of 2D corners of complementary chiral media can focus a source back on to itself, (b) The cylindrical coordinates, (c) A layered system
    with a periodic set of sources.}
    \label{figure6}
\end{figure}

Using a general method of coordinates mapping, one can map Maxwell-Tellgen's equations to other geometries and obtain a class of generalized perfect lenses in curvilinear coordinates. For instance, two negative two-dimensional corners, within which light radiating from a line source is bent around a closed trajectory and is refocused back onto itself \cite{notomi}, can be mapped to a layered structure consisting of layers of complementary bianisotropic media together with a periodic set of line sources, as shown in figure \ref{figure6}. Alternatively, the inverse transformation from a layered structure to 2D corners is available.

Here, the parameters of the chiral medium have been considered in the most general case which can take any values. Specifically, permittivity and permeability of a perfect lens proposed by Pendry is supposed to be -1, while for a perfect lens with chiral medium, its refractive index is $n=\sqrt{\ep \mu/\ep_0\mu_0} \pm c_0\xi$, that $c_0$ is the speed of light in vacuum and $\ep_0\mu_0=c_0^{-2}$, hence by taking proper values of the permittivity, permeability and chirality, we can obtain $n$, correspondingly, its complementary medium has $-n$, for a source placed in any one cell of the 2D corners, we have an image point in every cell of the corners.

Here, we would like to give a proof of the mapping from 2D corners onto periodic layered systems. For this, let us consider,
\begin{equation}
\begin{array}{cc}
  \varepsilon(\varphi)=+\varepsilon(\varphi), \quad \mu(\varphi)=+\mu(\varphi), \quad \xi(\varphi)=+\xi(\varphi), \quad \pi/2<\varphi<\pi \\[2mm]
  \varepsilon(\varphi)=-\varepsilon(\varphi), \quad \mu(\varphi)=-\mu(\varphi), \quad \xi(\varphi)=-\xi(\varphi), \quad 0<\varphi<\pi/2 \\[2mm]
\end{array}
\label{corner}
\end{equation}
then let us introduce a mapping of coordinates that takes structure in the left panel of figure \ref{figure6} into the structure of the right panel.
\begin{equation}
  x=r_0 \cos{\varphi}\, e^{l/l_0},\quad y=r_0 \sin{\varphi}\,e^{l/l_0},\quad z=Z
\end{equation}
while $r_0$, $\varphi$ and $Z$ are the cylindrical coordinates, in which the layered structure is defined. In this frame,
\begin{equation}
  \widetilde{\ep}_i=\ep_i\dfrac{Q_1Q_2Q_3}{Q_i^2},\quad \widetilde{\mu}_i=\mu_i\dfrac{Q_1Q_2Q_3}{Q_i^2},\quad \widetilde{\xi}_i=\xi_i\dfrac{Q_1Q_2Q_3}{Q_i^2}
\label{para}
\end{equation}
where,
\begin{equation}
  \begin{array}{l}
  Q_l=r_0/l_0 \sqrt{e^{2l/l_0}\cos^2{\varphi}+e^{2l/l_0}\sin^2{\varphi}}=r_0/l_0 e^{l/l_0} \\[2mm]
  Q_{\varphi}=r_0 \sqrt{e^{2l/l_0}\cos^2{\varphi}+e^{2l/l_0}\sin^2{\varphi}}=r_0 e^{l/l_0} \\[2mm]
  Q_Z=1 \\
  Q_l Q_{\varphi} Q_Z=r_0^2/l_0 e^{l/l_0}
  \end{array}
\end{equation}
and hence,
\begin{equation}
  \begin{array}{l}
    \tilde{\ep}_l=l_0 \ep_l, \quad \tilde{\ep}_{\varphi}=l_0^{-1} \ep_{\varphi},\quad \tilde{\ep}_Z=r_0^2/l_0 e^{2l/l_0}\ep_Z \\[2mm]
    \tilde{\mu}_l=l_0 \mu_l, \quad \tilde{\mu}_{\varphi}=l_0^{-1} \mu_{\varphi},\quad \tilde{\mu}_Z=r_0^2/l_0 e^{2l/l_0}\mu_Z \\[2mm]
    \tilde{\xi}_l=l_0 \xi_l, \quad \tilde{\xi}_{\varphi}=l_0^{-1} \xi_{\varphi},\quad \tilde{\xi}_Z=r_0^2/l_0 e^{2l/l_0}\xi_Z
  \end{array}
\end{equation}
substituting from (\ref{corner}) and setting $l_0=1$ gives,
\begin{equation}
  \begin{array}{l}
    \pi/2<\varphi<\pi: \\[2mm]
    \tilde{\ep}_l=\tilde{\ep}_{\varphi}=\ep(\varphi), \quad \tilde{\ep}_Z=r_0^2 e^{2l}\ep(\varphi) \\[2mm]
    \tilde{\mu}_l=\tilde{\mu}_{\varphi}=\mu(\varphi),\quad \tilde{\mu}_Z=r_0^2 e^{2l}\mu(\varphi) \\[2mm]
    \tilde{\xi}_l=\tilde{\xi}_{\varphi}=\xi(\varphi),\quad \tilde{\xi}_Z=r_0^2 e^{2l}\xi(\varphi)  \\[2mm]
    0<\varphi<\pi/2: \\[2mm]
    \tilde{\ep}_l=\tilde{\ep}_{\varphi}=-\ep(\varphi), \quad \tilde{\ep}_Z=-r_0^2 e^{2l}\ep(\varphi) \\[2mm]
    \tilde{\mu}_l=\tilde{\mu}_{\varphi}=-\mu(\varphi),\quad \tilde{\mu}_Z=-r_0^2 e^{2l}\mu(\varphi) \\[2mm]
    \tilde{\xi}_l=\tilde{\xi}_{\varphi}=-\xi(\varphi),\quad \tilde{\xi}_Z=-r_0^2 e^{2l}\xi(\varphi)
  \end{array}
\end{equation}
which can be indicated in the 1D periodic layers with an alternation of complementary bianisotropic medium.

\subsection{From 2D checkerboards to 3D corner lenses}
\begin{figure}
    \centering
    \includegraphics[scale=0.6]{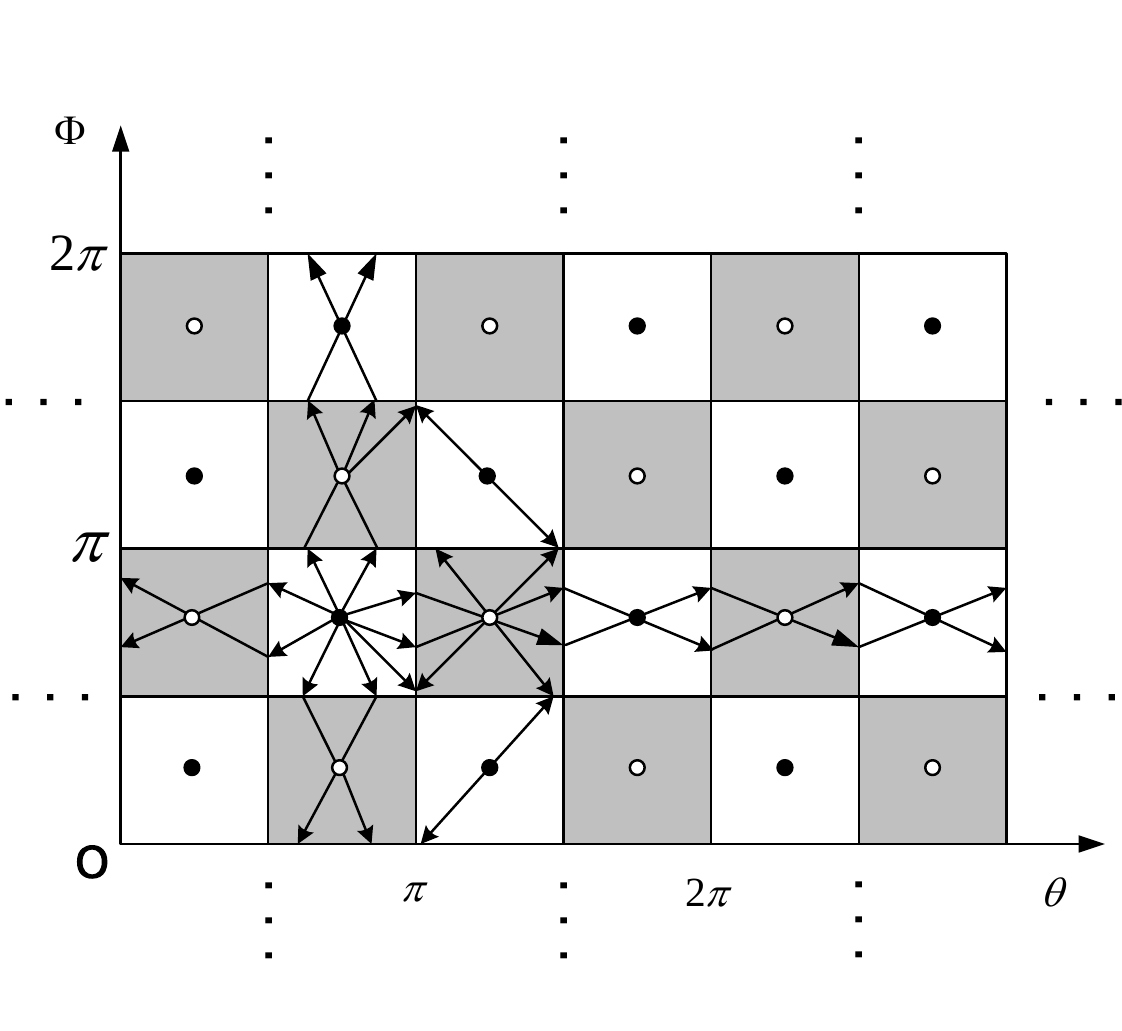}
    \caption{An infinite checkerboard with optically complementary cells. When there is a source in one cell of the checkerboard, then according to the ray routes, images are recreated in every other cell.}
    \label{figure7}
\end{figure}
Figure \ref{figure7} shows a 2D checkerboard consisting an alteration of rectangular blocks of complementary bianisotropic media. The axes along the checkerboard directions are defined by $\theta$ and $\phi$, along which the periodic of the checkerboard is $\pi$, the direction normal to this plane is denoted by $l$. The structure is defined,
\begin{equation}
  \begin{array}{l}
  \ep(\theta,\phi)=+\ep(\theta,\phi), \,\, \mu(\theta,\phi)=+\mu(\theta,\phi), \,\, \xi(\theta,\phi)=+\xi(\theta,\phi), \,\, \forall \, (\theta,\phi)\in {\rm positive}, \\[2mm]
  \ep(\theta,\phi)=-\ep(\theta,\phi), \,\, \mu(\theta,\phi)=-\mu(\theta,\phi), \,\, \xi(\theta,\phi)=-\xi(\theta,\phi), \,\, \forall \, (\theta,\phi)\in {\rm negative},
  \end{array}
\end{equation}
where
\begin{equation}
  \begin{array}{c}
    {\rm positive}=(0,\pi/2)\times (0,\pi/2)\cup(\pi/2,\pi)\times(\pi/2,\pi), \\[2mm]
    {\rm negative}=(0,\pi/2)\times (\pi/2,\pi)\cup(\pi/2,\pi)\times(0,\pi/2),
  \end{array}
\end{equation}
Obviously, a refocussing can be observed in each cell while the rays starting from a single source in a positive cell as shown in figure \ref{figure7}. This phenomenon can be unveiled by the generalized lens theorem, let us consider the imaging along the $\theta$ direction, in which direction the condition of complementarity is satisfied for the layers, hence, for a source at $\theta=\theta_1$, $\phi=\phi_1$ in a positive cell, we can obtain a set of images along $\phi=\phi_1$
line at $\theta=\pm(n\pi+\theta_1)$, with $n$ a positive integer. Similarly, a set of images along $\phi$ direction can be obtained, and finally an image in every cell of the checkerboard can be achieved corresponding to the source placed in any one cell.
\begin{figure}
    \centering
    \includegraphics[scale=0.65]{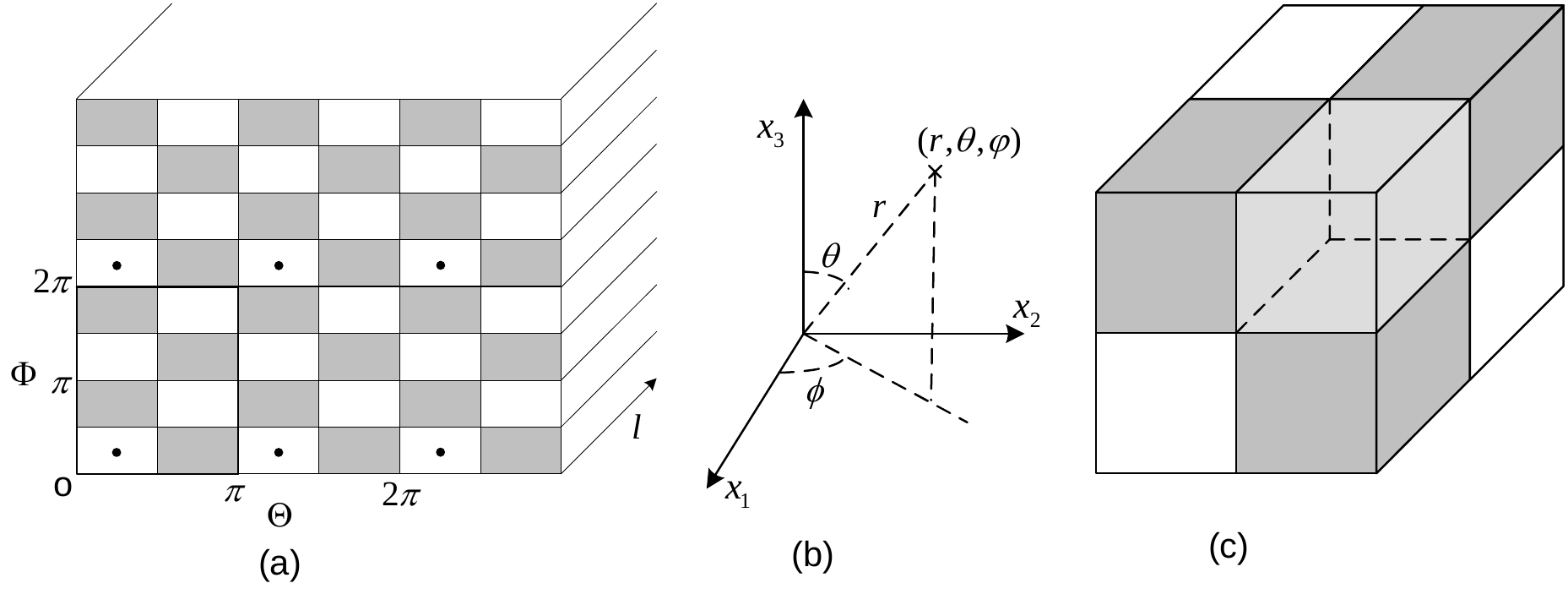}
    \caption{(a) 2D checkerboard with optically complementary cells.(b) The spherical coordinates.(c)Cubic corner lens.}
    \label{figure8}
\end{figure}

According to proper coordinates transformation, a 2D corner can be mapped to a 1D lens which satisfied the complementary conditions. Furthermore, let us consider the mapping algorithm from a 2D checkerboard to 3D cubic corner as shown in figure \ref{figure8}. As shown in the left panel of figure \ref{figure8}, the checkerboard consisting of a doubly periodic set of point sources with period $2\pi$ along $\phi$ and period $\pi$ along $\theta$, the coordinates of which is denoted by $[l,\theta,\phi]$; the right panel is corresponding to a 3D cubic corner extending to infinity, of which the coordinates are denoted by Cartesian coordinates $[x_1,x_2,x_3]$. According to the spherical coordinate system indicated in the middle panel, the two coordinates satisfy,
\begin{equation}
  \begin{array}{l}
  l=\dfrac{l_0}{2}{\rm ln}(\dfrac{x_1^2+x_2^2+x_3^2}{r_0^2}),\\[3mm]
  \theta=\arccos(\dfrac{x_3}{\sqrt{x_1^2+x_2^2+x_3^2}}),\\[5mm]
  \phi=\arctan(\dfrac{x_2}{x_1})
  \end{array}
\end{equation}
where $r_0$ is a scale factor and can be considered as the radial position of the source. $l$ denotes the radial (logarithmic) coordinate and $\theta$, $\phi$ are the longitudinal and azimuthal coordinates; we can then generate the corresponding cubic corner system in $[x_1, x_2,x_3]$.

Considering a checkerboard with homogeneous and isotropic cells, then according to the coordinates transformation algorithm, we have
\begin{equation}
  \begin{array}{l}
    Q_1=\sqrt{\dfrac{l_0^2r_0^4x_1^2}{r^4}+\dfrac{x_1^2x_3^2}{r^4(x_1^2+x_2^2)}+\dfrac{x_2^2}{(x_1^2+x_2^2)^2}}, \\[3mm]
    Q_2=\sqrt{\dfrac{l_0^2r_0^4x_2^2}{r^4}+\dfrac{x_2^2x_3^2}{r^4(x_1^2+x_2^2)}+\dfrac{x_1^2}{(x_1^2+x_2^2)^2}}, \\[3mm]
    Q_3=\sqrt{\dfrac{l_0^2r_0^4x_3^2}{r^4}+\dfrac{x_1^2+x_2^2}{r^4}},
  \end{array}
\label{Q}
\end{equation}
with $r=\sqrt{x_1^2+x_2^2+x_3^2}$. Assuming $l_0=1$, $r_0=1$ and
\begin{equation}
  x_1=r\sin{\theta}\cos{\phi},\quad x_2=r\sin{\theta}\sin{\phi}, \quad x_3=r\cos{\theta},
\end{equation}
(\ref{Q}) can be recasted as,
\begin{equation}
  Q_1=\dfrac{1}{r}\sqrt{\cos^2{\phi}+\dfrac{\sin^2{\phi}}{\sin^2{\theta}}},\quad
  Q_2=\dfrac{1}{r}\sqrt{\sin^2{\phi}+\dfrac{\cos^2{\phi}}{\sin^2{\theta}}},\quad Q_3=\dfrac{1}{r}.
\label{Q2}
\end{equation}
on the other hand, we have (\ref{para}) as the new set of the parameters, hence let us substitute (\ref{Q2}) into (\ref{para}), we can deduce
\begin{equation}
  \begin{array}{l}
    \widetilde{v}_1=\dfrac{v_1}{r}(\dfrac{\sin^2{\theta}\sin^2{\phi}+\cos^2{\phi}}{\sin^2{\theta}\cos^2{\phi}+\sin^2{\phi}}),\\[3mm]
    \widetilde{v}_2=\dfrac{v_2}{r}(\dfrac{\sin^2{\theta}\cos^2{\phi}+\sin^2{\phi}}{\sin^2{\theta}\sin^2{\phi}+\cos^2{\phi}}),\\[3mm]
    \widetilde{v}_3=\dfrac{v_3}{r\sin^2{\theta}}\sqrt{\sin^2{\theta}\cos^2{\phi}+\sin^2{\phi}}\sqrt{\sin^2{\theta}\sin^2{\phi}+\cos^2{\phi}}
  \end{array}
\end{equation}
where $v=\ep, \mu,\xi$. It is noted that the transverse $r$ is irrelevant to the imaging. Now, choosing
\begin{equation}
  \begin{array}{l}
  \ep_1=\mu_1=\xi_1=r(\dfrac{\sin^2{\theta}\cos^2{\phi}+\sin^2{\phi}}{\sin^2{\theta}\sin^2{\phi}+\cos^2{\phi}}),\\[3mm]
  \ep_2=\ep_1^{-1}=\mu_2=\mu_1^{-1}=\xi_2=\ep_1^{-1}, \\[2mm]
  \ep_3=\mu_3=\xi_3=r\sin^2{\theta}\dfrac{1}{\sqrt{\sin^2{\theta}\cos^2{\phi}+\sin^2{\phi}}}\dfrac{1}{\sqrt{\cos^2{\theta}\sin^2{\phi}+\cos^2{\phi}}}
  \end{array}
\end{equation}
then we obtain that $(\widetilde{\ep}_1,\widetilde{\ep}_2,\widetilde{\ep}_3)$, $(\widetilde{\mu}_1,\widetilde{\mu}_2,\widetilde{\mu}_3)$ and $(\widetilde{\xi}_1,\widetilde{\xi}_2,\widetilde{\xi}_3)$ to be identical to the checkerboard made of homogeneous, isotropic materials. So far, we have proved that we can achieve a cubical corner lens made up of anisotropic, inhomogeneous materials by mapping from a homogeneous checkerboard.


\section{Numerical illustration of finite element method}
\subsection{Diffraction problem for fully bianisotropic media}
In this section, we would like to give a numerical illustration to the structures consisting of complementary bianisotropic media through FEM (implemented in COMSOL MULTIPHYSICS). First, we would like to derive the diffraction equations for a bianisotropic medium, generally, we suppose those parameters in equation (\ref{max}), such as permittivity, permeability and chirality are all tensors written as follows,
\begin{equation}
\varepsilon =\left[\begin{array}{ccc}
\varepsilon_{11} & \varepsilon_{12} & 0\\
\varepsilon_{21} & \varepsilon_{22} & 0\\
0 & 0 & \varepsilon_{33}
\end{array}\right] , \quad
\mu =\left[\begin{array}{ccc}
\mu_{11} & \mu_{12} & 0\\
\mu_{21} & \mu_{22} & 0\\
0 & 0 & \mu_{33}
\end{array}\right] , \quad
\xi =\left[\begin{array}{ccc}
\xi_{11} & \xi_{12} & 0\\
\xi_{21} & \xi_{22} & 0\\
0 & 0 & \xi_{33}
\end{array}\right]
\label{f10}
\end{equation}
and also assume that ${\bf E} = \left(e_1,e_2,e\right)^T$,${\bf H} = \left(h_1,h_2,h\right)^T$,
which leads to
\begin{equation}
\begin{array}{ccc}
\displaystyle{\nabla \cdot \left\{(\underline{\underline{\mu_T}}+\underline{\underline{\xi_T}} \; {\underline{\underline{\varepsilon_T}}}^{-1} \;\underline{\underline{\xi_T}})^{-1}
\nabla e \right\}
+i \nabla \cdot \left\{(\underline{\underline{\xi_T}} + \underline{\underline{\varepsilon_T}}\; {\underline{\underline{\xi_T}}}^{-1}\; \underline{\underline{\mu_T}})^{-1}
\nabla h \right\} }\\[2mm]
=\omega^2 \varepsilon_{33}e + i\omega^2 \xi_{33}h  \\[2mm]
i\displaystyle{\nabla \cdot \left\{(\underline{\underline{\xi_T}}^T + \underline{\underline{\mu_T}}\; {\underline{\underline{\xi_T}}}^{-1} \;\underline{\underline{\varepsilon_T}})^{-1}
\nabla e \right\}
+\nabla \cdot \left\{(\underline{\underline{\varepsilon_T}}+\underline{\underline{\xi_T}}\; {\underline{\underline{\mu_T}}}^{-1}\; \underline{\underline{\xi_T}}^T)^{-1}
\nabla h \right\} } \\[2mm]
=i\omega^2 \xi_{33}e - \omega^2 \mu_{33}h\\
\end{array}
\label{f15}
\end{equation}
with
\begin{equation}
\underline{\underline{\varepsilon_T}}=\left[\begin{array}{cc}
\varepsilon_{22} & -\varepsilon_{21}\\
-\varepsilon_{12} & \varepsilon_{11}
\end{array}\right] ,\quad
\underline{\underline{\mu_T}}=\left[\begin{array}{cc}
-\mu_{22} & \mu_{21}\\
\mu_{12} & -\mu_{11}
\end{array}\right] ,\quad
\underline{\underline{\xi_T}}=\left[\begin{array}{cc}
-\xi_{22} & \xi_{21}\\
\xi_{12} & -\xi_{11}
\end{array}\right]
\label{f14}
\end{equation}
Based on this, we can set up any proposed model indicated by equation (\ref{f15}) in COMSOL MULTIPHYSICS by using PDE module.
We note that such a model is required e.g. in order to implement perfectly matched layers (for which $\varepsilon$, $\mu$
and $\xi$ are all tensors). In the sequel, we apply this numerical model to various types of complementary media.

\subsection{Complementary media with isotropic $\varepsilon$, $\mu$ and $\xi$}
\begin{figure}
    \centering
    \includegraphics[scale=0.6]{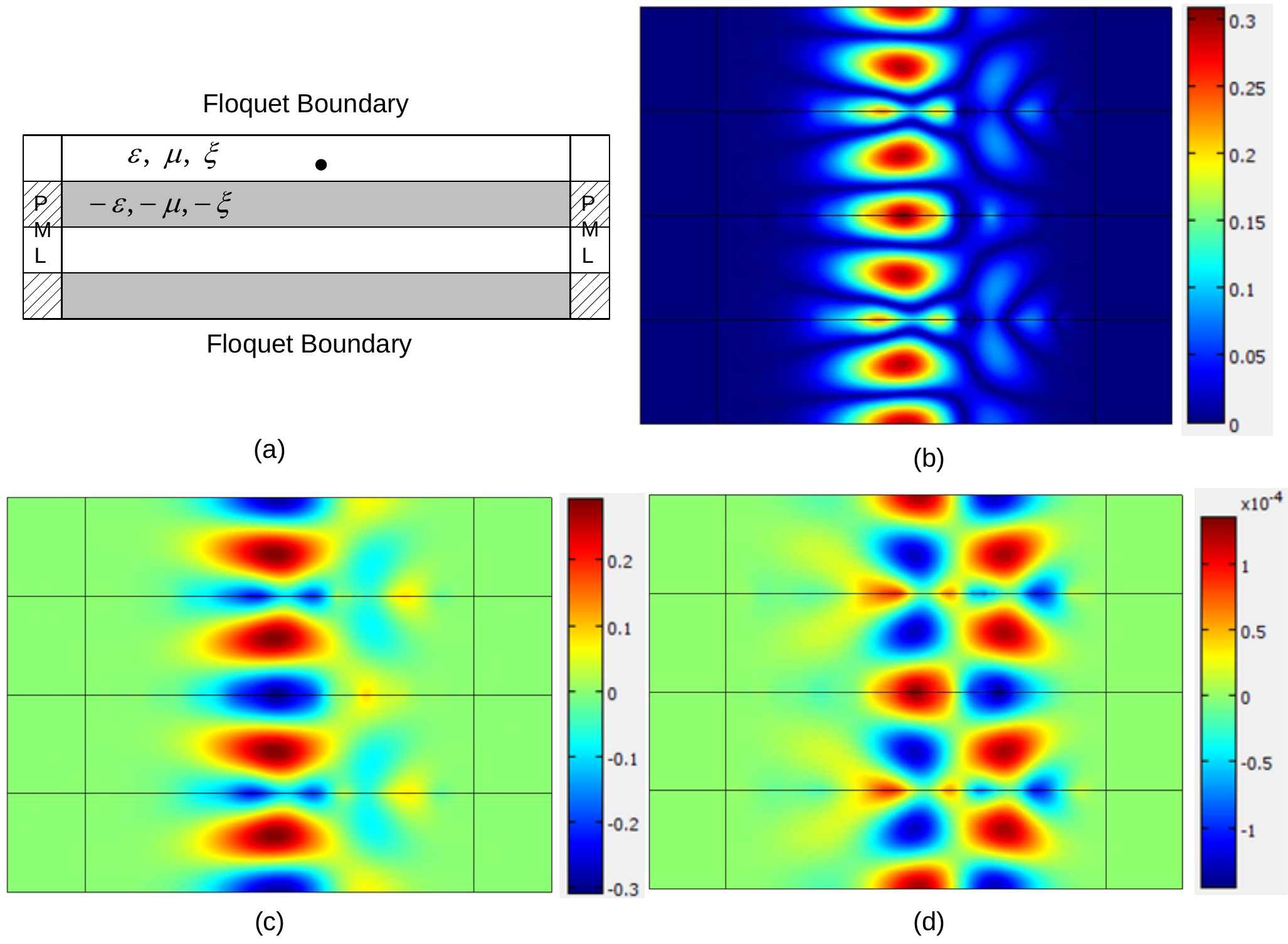}
    \caption{Periodic 1D complementary bianisotropic media: (a) Schematic view of the computational model with Perfectly Matched Layers (PMLs) on either sides of the structure, periodic conditions on top and bottom and a line source in the top layer; (b) 2D plot of $\sqrt{e^2+h^2}$ for a time harmonic electric line source radiating in a periodic set of complementary chiral lenses; (c) Plot of real part of electric field $e$; (d) Plot of real part of magnetic field $h$.}
    \label{figure9}
\end{figure}

In this part, we would like to give a numerical analysis to the structure of complementary bianisotropic media described by isotropic (sign-shifting) parameters $\varepsilon$, $\mu$
and $\xi$. First, we analyze an infinite chiral lens, which consists of the complementary bianisotropic medium as shown in the left panel of figure \ref{figure9}, while there is a line source lying in the top bianisotropic layer,
in order to asymptotic the infinite half space, we have introduced the Floquet-Bloch boundary at the upper and lower interface, and perfect matched layers at the left and right sides of the structure. The FEM results is shown in the right panel, a refocussing of the source can be observed in each layer as indicted in previous theoretical analysis. Here, for the module in COMSOL, we have supposed $\ep=\ep_0$, $\mu=\mu_0$ and $\xi/c_0=0.99$ respectively, with the normalized frequency of the source is $f=7\times10^{14}{\rm Hz}$.

\begin{figure}
    \centering
    \includegraphics[scale=0.6]{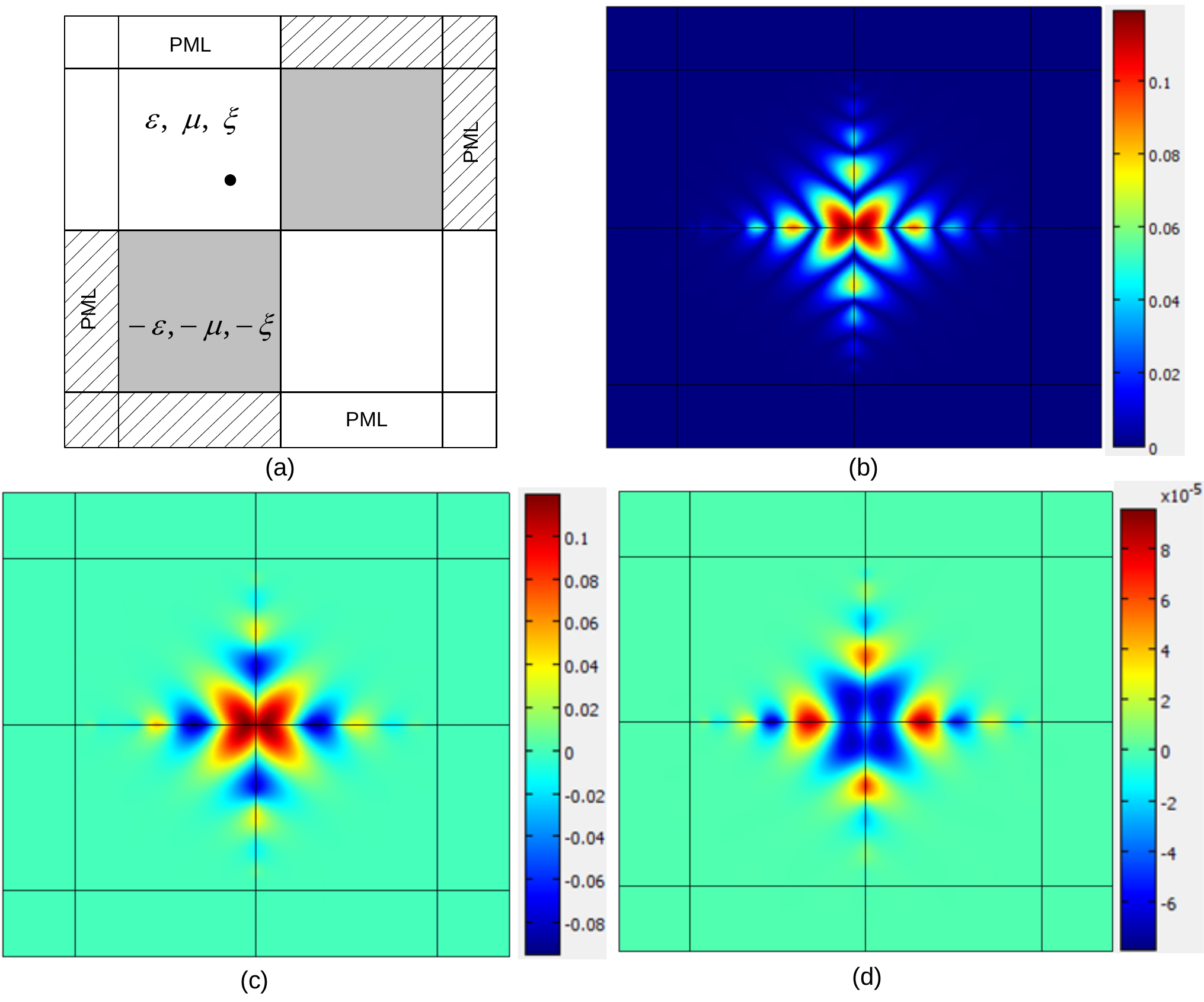}
    \caption{2D complementary bianisotropic corner: (a) Schematic view of the computational model with Perfectly Matched Layers (PMLs) on either sides of the structure, and a line source in one of the corner; (b) 2D plot of $\sqrt{e^2+h^2}$ for a time harmonic electric line source radiating in a 2D complementary bianisotropic corners;(c) Plot of real part of electric field $e$; (d) Plot of real part of magnetic field $h$.}
    \label{figure10}
\end{figure}

Furthermore, a 2D chiral corner is also considered as shown in figure \ref{figure10} (the size of structure is in order of $um$), with a line source lying in one of corner with positive parameters, similarly, the source has been mapped to the other corners as an image. The parameters of the media are supposed to be $\ep=\ep_0$, $\mu=\mu_0$ and $\xi/c_0=0.99$ respectively, where the normalized frequency of the source is $f=6\times10^{14}{\rm Hz}$.

\begin{figure}
    \centering
    \includegraphics[scale=0.6]{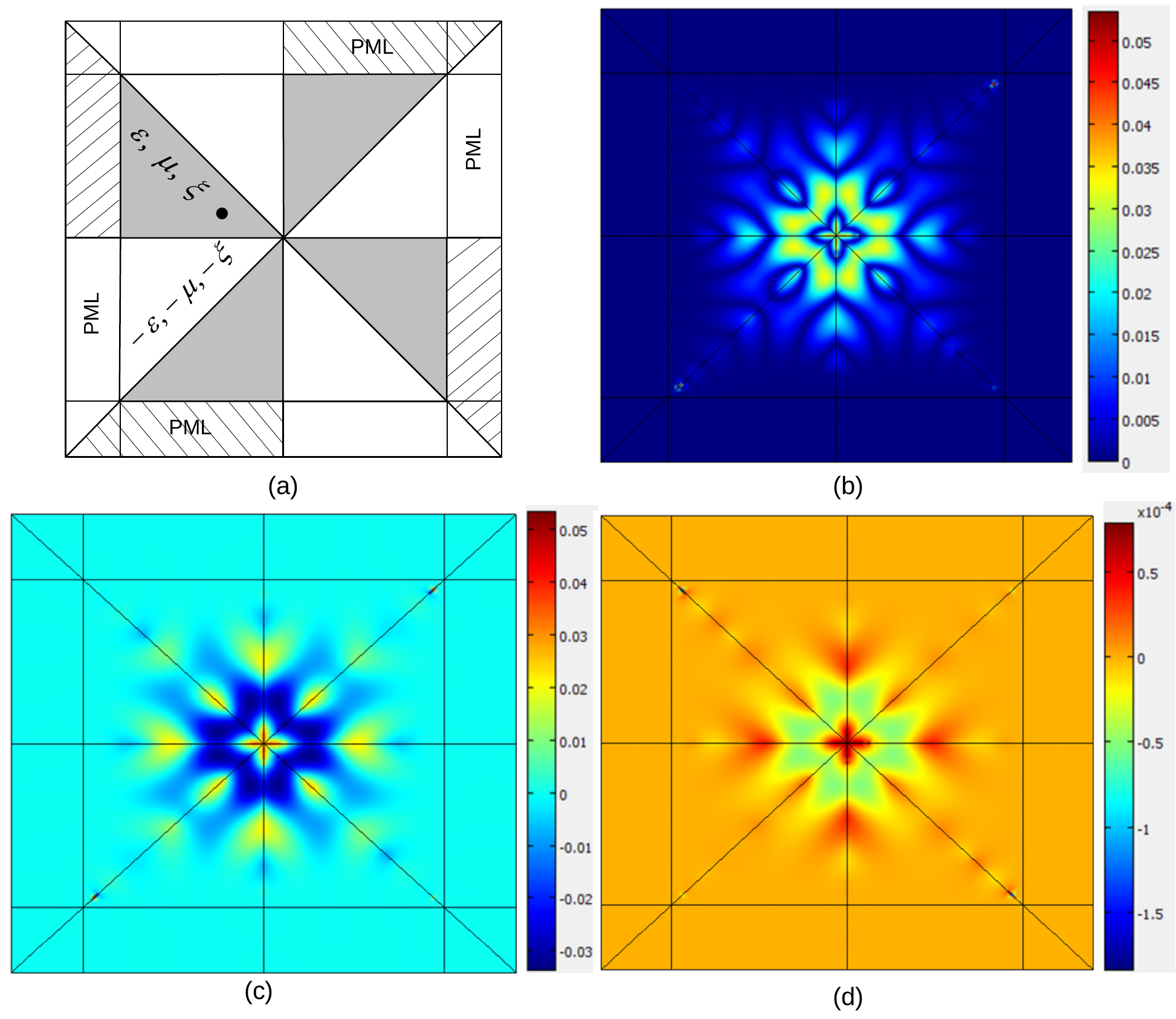}
    \caption{2D complementary bianisotropic corner with an alteration of triangle shape: (a) Schematic view of the computational model with Perfectly Matched Layers (PMLs) on either sides of the structure, and bottom and a line source in the top layer; (b) 2D plot of $\sqrt{e^2+h^2}$ for a time harmonic electric line source radiating in a periodic set of complementary bianisotropic corners;(c) Plot of real part of electric field $e$; (d) Plot of real part of magnetic field $h$.}
    \label{figure11}
\end{figure}

In the same way, a triangle corner lens as shown in figure \ref{figure11} is simulated, with $\ep=\ep_0$, $\mu=\mu_0$ and $\xi/c_0=0.99$ respectively, and the normalized frequency of the source is $f=7\times10^{14}{\rm Hz}$.

\begin{figure}
    \centering
    \includegraphics[scale=0.7]{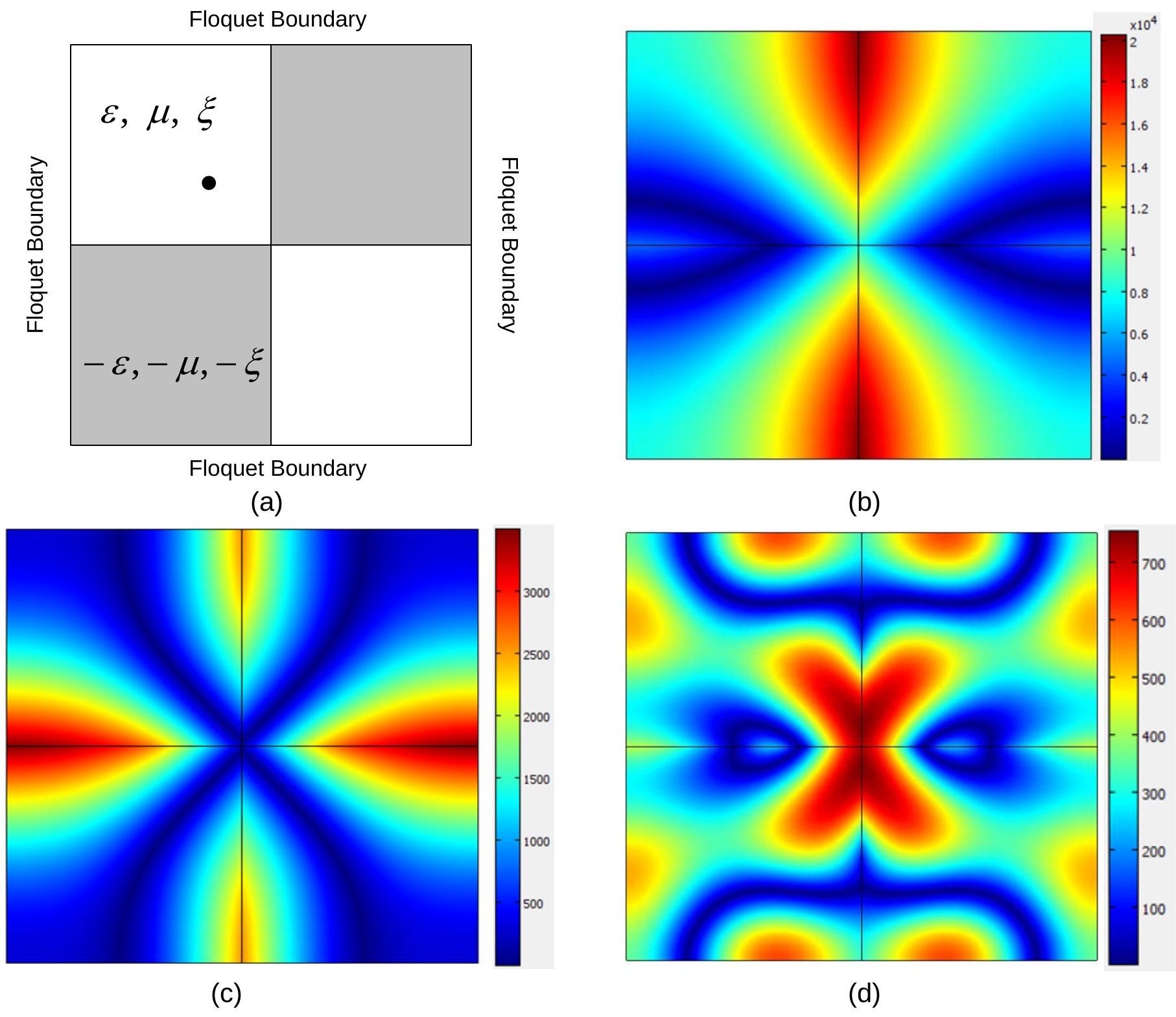}
    \caption{An infinite checkerboard consisting with complementary bianisotropic media:(a) Schematic view of the computational model with Floquet Boundary conditions in either side of the unit cell; (b)-(d) Plots of $\sqrt{e^2+h^2}$ for a line source radiating in an infinite checkerboard of complementary bianisotropic media.}
    \label{figure12}
\end{figure}

At last, a checkerboard and its numerical results by COMSOL are shown in figure \ref{figure12}, the permittivity and permeability are $\ep=\ep_0$, $\mu=\mu_0$, while the chirality is $\xi/c_0=0.99$, the frequency is $f=3\times10^{14}{\rm Hz}$.

One should note that absence of absorption in sign-shifting bianisotropic media means the Maxwell-Tellegen operator is no longer elliptic and hence Lax-Milgram¡¯s lemma ensuring existence and uniqueness of the solution to the diffraction problem is no longer applicable. Nevertheless, the Fredholm alternative ensures that if there is a solution for any applied electromagnetic field whose tangential component is continuous through interfaces separating complementary media (hence exhibiting two anti-parallel wave-vectors at both sides of such interfaces) and satisfying periodic conditions on the unit cell edges, it will be an eigenfunction for the problem as shown in figure \ref{figure12}(b)-(d). The convergence of the Finite Element algorithm depends crucially upon the regularity of the boundaries between complementary media with some ill-posedness occurring in certain corner cases for a refractive index contrast of $-1$ when the coercive plus compact mathematical framework breaks down, see \cite{Bonnet-Ben Dhia} for the case when $\xi=0$. Thus, our computer calculations represent a benchmark for validation of numerical results involving bianisotropic materials in very singular conditions such as those presented here.

\subsection{Numerical results for anisotropic $\varepsilon$, $\mu$ and $\xi$}
Let us now look at the case of a cylindrical lens with tensors of permittivity, permeability and chirality
as follows
\begin{equation}
  v_I={\rm diag}[v_x^{(I)},\,v_y^{(I)},\,v_z^{(I)}], \quad v=\ep, \, \mu,\, \xi
\end{equation}
with the index $I=1,2,3$ corresponding to the three regions of the cylindrical lens
\begin{equation}
\begin{array}{l}
  v^{(1)}_x=+1, \, v^{(1)}_y=+1, \, v^{(1)}_z=+1, \, r \leq r_1 \\[2mm]
  v^{(2)}_x=-1, \, v^{(2)}_y=-1, \, v^{(2)}_z=-r_2^4/r^4, \, r_1< r \leq r_2 \\[2mm]
  v^{(3)}_x=+1, \, v^{(3)}_y=+1, \, v^{(3)}_z=r_2^4/r_1^4, \, r_2 < r \leq r_3
\end{array}
\end{equation}
where region 3 is the vacuum surrounding the cylindrical lens, region 2 is filled with negative bianisotropic media, which is designed to optically cancel out region 3 ($r_3=r^2_2/r_1$), and region 1 is interior part of lens; the whole structure acts that the region 1 is optically equivalent to vacuum of radius $r_3$. Hence, for an object lying in region 1, its image will be magnified by a factor of $r_3/r_1=r_2^2/r_3^2$ in region 3.

\begin{figure}
    \centering
    \includegraphics[scale=0.65]{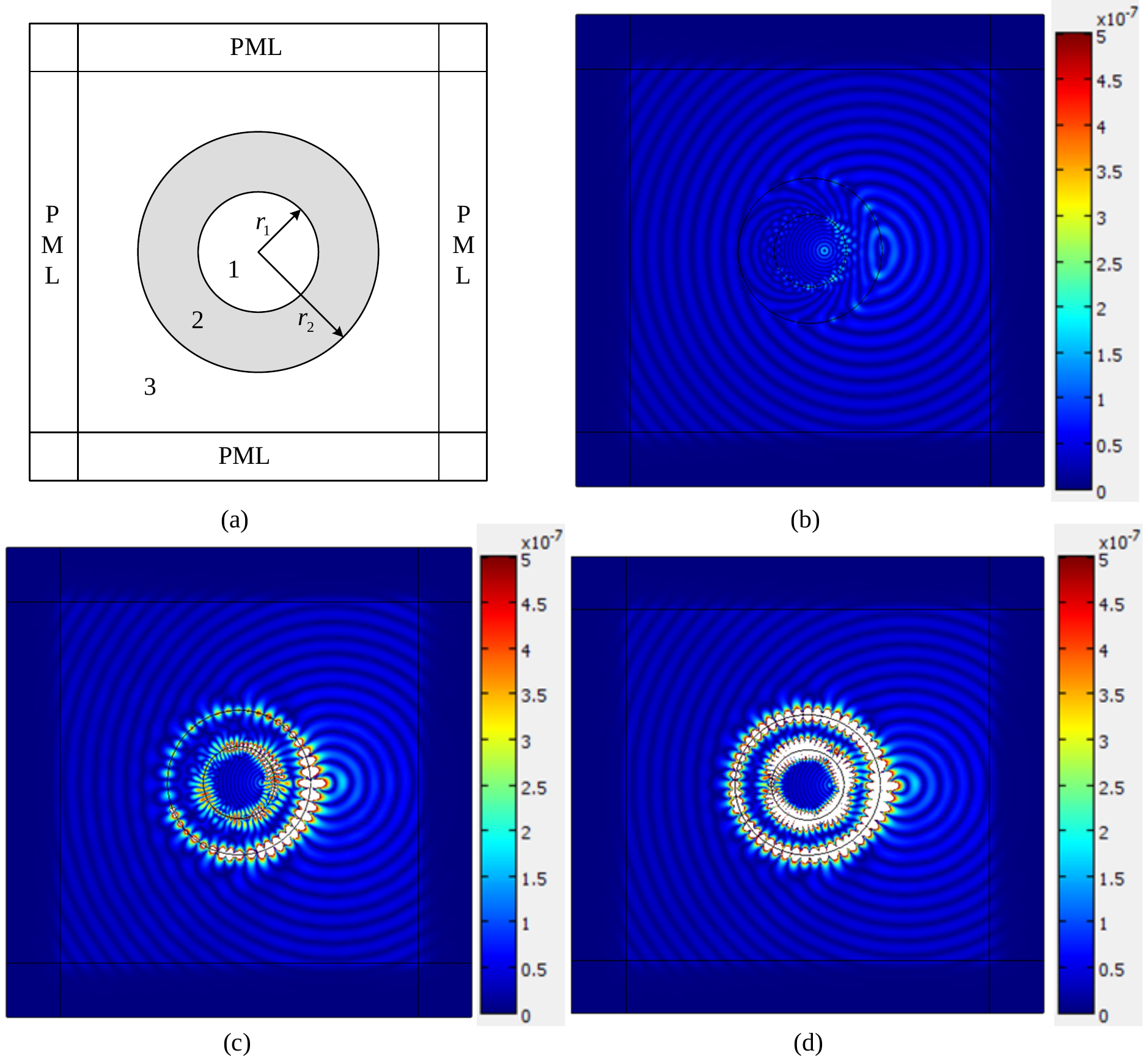}
    \caption{A cylindrical lens consisting with complementary bianisotropic media:(a) Schematic view of the computational model with PML in either side of the unit cell; (b) Plots of $\sqrt{e^2+h^2}$ for a line source radiating in region 1 and $r>r_1^2/r_2$; (c)Plots of $\sqrt{e^2+h^2}$ for a line source radiating in region 1 and $r<r_1^2/r_2$;(d)Plots of $\sqrt{e^2+h^2}$ for a line source radiating in the annulus with $r=r_1+(r_2-r_1)/2$ along the horizontal axis from the origin.}
    \label{figure13}
\end{figure}

It should be noted that when the source placed at $r>r_1^2/r_2$, images will be formed in both the annulus and the outer vacuum space as shown in figure \ref{figure13}(b); otherwise, when  $r<r_1^2/r_2$, only an image will be formed in region 3 in figure \ref{figure13} (c); while when a source is placed in the annulus, then there will be two images respectively in region 1 and 3 as in figure \ref{figure13} (d).
where frequency $f=8.7\times10^{14}{\rm Hz}$, and the radius $r_1=0.2um$, $r_2=0.4um$.

It is interesting to note that anomalous resonances on either sides of the cylindrical lens crucially depend upon absorption within the shell, see \cite{milton1}, and we have checked that introducing some small imaginary part to one component of the permittivity, permeability or chiral tensor is enough to remove the singularities on the shell's boundaries. These anomalous resonances can be used in a way similar to what was done in \cite{milton2} to cloak some dipole sources outside a shell with negative permittivty, but this lies beyond the scope of the present paper.

\section{Concluding remarks}
In conclusion, we have derived a generalized perfect theorem that includes bianisotropic and chiral media in addition to the usual dielectric and magnetic media. We proposed two different methods to derive a generalized lens theorem for isotropic (section 2.1) and anisotropic (section 2.2), sign-shifting permittivity, permeability and chirality admittance tensors. The geometric transformation technique (section 3.1) is generalized to include bianisotropic media as well and the two results are applied to analyze a variety of singular situations involving corners and wedges of such media. These results are numerically simulated by finite element calculations using the COMSOL package and represent a benchmark for the validation of computations involving bianisotropic media in extremely singular circumstances.

\end{document}